\documentclass[twocolumn, amsmath,amssymb, superscriptaddress]{revtex4}

\usepackage{dcolumn}
\usepackage{bm}
\usepackage{amsmath}
\usepackage{amsfonts}
\usepackage{graphics}
\usepackage[dvips]{graphicx}
\usepackage{times}
\usepackage{setspace}
\usepackage{color}   
\usepackage{verbatim}
\usepackage[raggedright]{titlesec}
\usepackage[normalem]{ulem}
\usepackage{framed}

\begin{document}
\title{ A quantitative perspective on ethics  in large team science}
\author{Alexander M. Petersen}
\affiliation{Laboratory for the Analysis of Complex Economic Systems, IMT Lucca Institute for Advanced Studies, Lucca 55100, Italy}
\author{Ioannis Pavlidis}
\affiliation{Computational Physiology Laboratory, University of Houston, Houston, Texas 77204}
\author{Ioanna Semendeferi}
\affiliation{College of Natural Sciences and Mathematics, University of Houston, Houston 77204, USA}

\begin{abstract} 
The gradual crowding out of singleton and small team science by large team endeavors is challenging key features of research culture. It is therefore important for the future of scientific practice to  reflect upon the individual scientist's ethical responsibilities within teams. To facilitate this reflection we show labor force trends in the US revealing a skewed growth in academic ranks and increased levels of competition for promotion within the system; we analyze teaming trends across disciplines and national borders demonstrating why it is becoming difficult to distribute credit and to avoid conflicts of  interest; and we use more than a century of Nobel prize data to show how science is outgrowing its old institutions of singleton awards. Of particular concern within the large team  environment is the weakening of the mentor-mentee relation, which undermines the cultivation of virtue ethics across scientific generations. These trends and emerging organizational complexities call for a universal set of behavioral norms that transcend team heterogeneity and hierarchy. To this end, our expository analysis provides a survey of ethical issues in team settings to inform science ethics education and science policy.
\end{abstract}

\maketitle

\footnotetext[1]{ Send correspondence to:  petersen.xander@gmail.com}

Many of science's grand challenges have become too daunting for individual investigators to undertake. The increase in the characteristic size and complexity of teams reflects the  division of labor that is necessary in large projects.  As a result,  team science is now more prevalent than individual science, a shift that has occurred slowly but steadily over the last century \citep{TeamScience,TeamFormationEvol}. 

The range in the size of scientific endeavors spans three orders of magnitude,  from singleton  to ``Big Science''  programs in excess of 1000 members \citep{LittleBigScience}. 
Large-scale multi-disciplinary projects,  requiring extensive resources,  have become increasingly common. Examples include the Higgs particle experiment at CERN, the big data genomics   project   by the ENCODE consortium \citep{ENCODE}, cross-institutional medical trials  \citep{OASIS5}, and  large scale  digital humanities projects such as the  Google Inc. n-gram portal \citep{michels11}. A better understanding of  team science is important for the economics of science \citep{StephanJEL,EconScience,OpenScience}, the management of science \citep{TeamScience,TeamAssembly,MultilevelScience}, the evaluation of scientific careers  \citep{GrowthCareers,EuropeTenureSystems,CollabMetrics}, and the  internationalization of science 
\citep{ScienceEurope,SantoWorldCollab}.

An open discussion focused on ethical issues germane to team science is also important for the future of scientific research, which ultimately depends on the quality of individual contributions.
In the academic domain, production of public knowledge is based upon priority, a type of credit that  incentivizes scientists to share, reuse, and build upon the knowledge stock \citep{StephanJEL,EconScience,OpenScience}. Two key features of this credit system are that the priority be clearly assignable and the credit be transparently divisible among coauthors. However, with increasing team size, typically accompanied by a hierarchical management structure, it has become  difficult to monitor and evaluate individual efforts towards team objectives, rendering a fair division of credit challenging \citep{CreditTaxonomy}. 

Here we take a quantitative historical approach \citep{CompHistorySci}  to initiate discourse on a class of ethical considerations that have emerged with team science and are in contradistinction to ethical guidelines in singleton science \citep{OnBeingAScientistShort,OnBeingAScientistLong}. These considerations are inherently complex because they span multiple levels of context, from the individual, to the team, and even up to the international level.

In what follows, we start in the Results section with empirical evidence that provides reference points in  the Discussion section. Specifically, we start by analyzing diverse data sources to document the skewed growth of the scientific labor force, the growth of  team size in science, the implications of large team size on hierarchy and transparency, the limits of individual achievement awards, such as the Nobel Prize, and the internationalization of scientific collaboration networks. We then draw upon  these quantitative illustrations in our  discussion of six ethics issues in team science: (i) the ethics of credit; (ii) the ethics of coauthorship; (iii) the ethical dilemmas associated with conflict of interest; (iv) the attenuation of the mentor-mentee relationship and the threat it poses to virtue ethics; (v) the ethical dilemmas manifesting in cross-border collaboration; and (vi) the universality of norms. \\

\begin{figure*}
\centering{\includegraphics[width=0.8\textwidth]{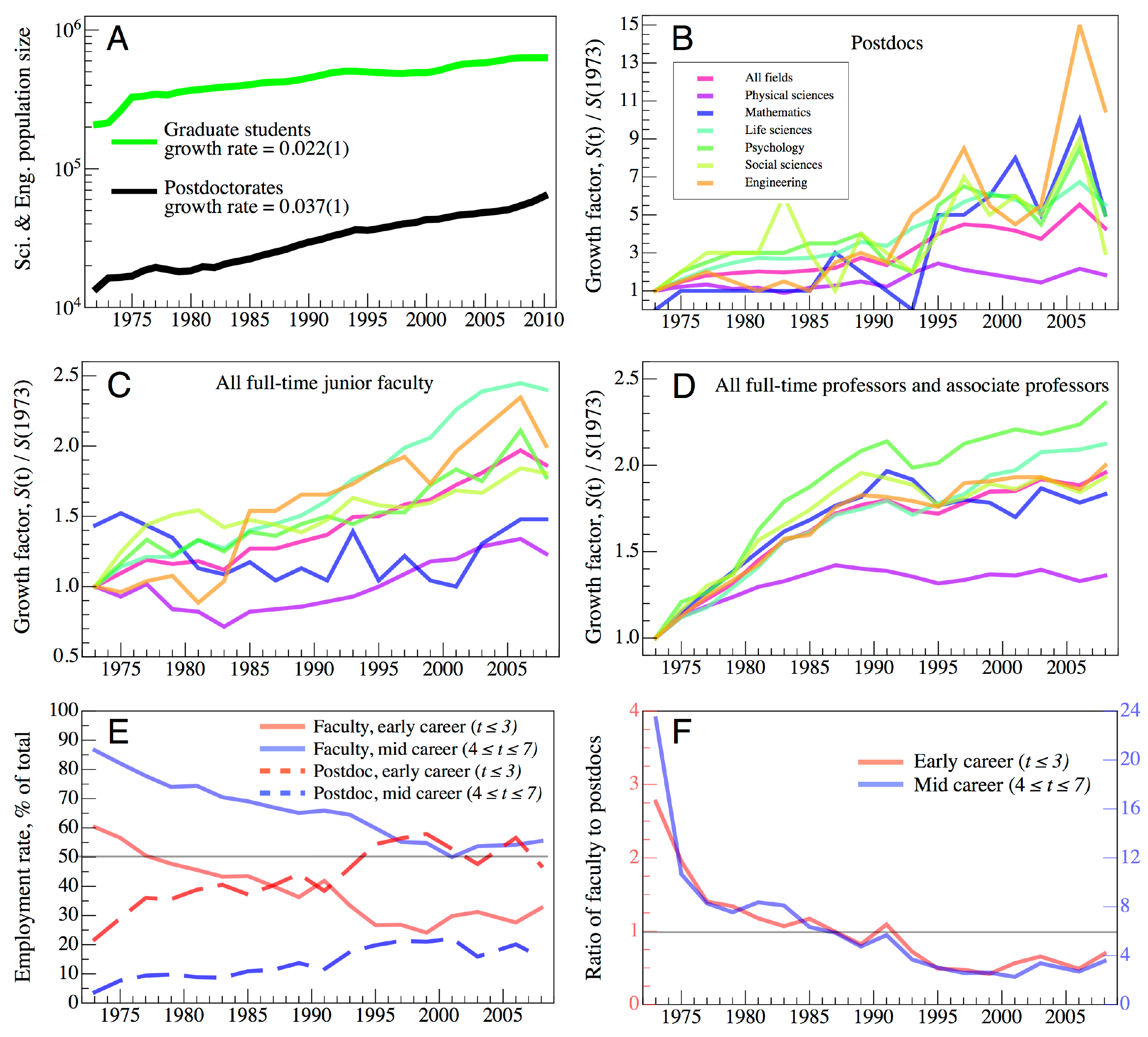}}
\caption{\label{f0}   {\bf Growth of the scientific labor force.} 
(A) The annual numbers of graduate students and postdoctorates in  US Science \& Engineering departments \citep{D0}.
(B,C,D) The annual numbers of US  postdoc and faculty positions by degree field \citep{D1}.
(E,F) The annual percentages of  doctorate holders by group (faculty vs. postdocs) and their ratios  in US universities. Data was aggregated over two distinct age cohorts: careers with 1-3 years (early) and 4-7 years (mid) since doctorate  \citep{D2}. The disproportionate growth rates between the faculty and postdoc positions may pose a threat to the mentor-mentee relationship. 
}
\end{figure*}

\begin{figure*}
\begin{center}
\includegraphics[width=0.8\textwidth]{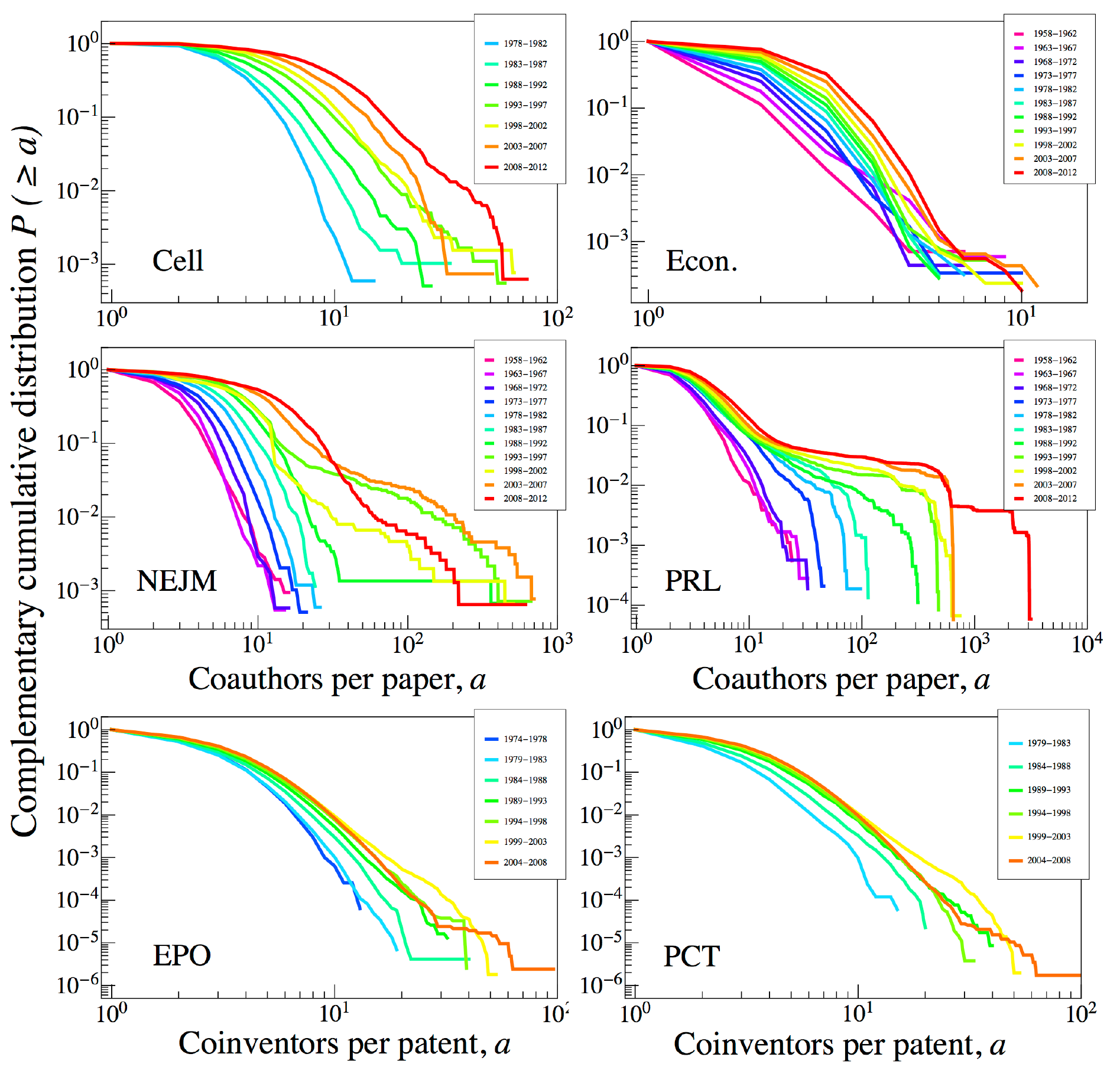}
\caption{\label{f1}  {\bf Expansion of the team-size distribution.} 
The observed frequency  $P(\geq a)$ of papers or patents  with team size of at least size $a$. The plots are shown on log-log axes, where each colored curve corresponds to   a non-overlapping 5-year period indicated in the legend. The broad distribution of $a$ values for each journal demonstrates that the credit for a single publication can be distributed across a far-ranging number of contributors, whereas for patents, the typical team size and its spread are considerably smaller.}
\end{center}
\end{figure*}

\section*{Results}
\noindent{\bf Skewed Growth of the Scientific Labor Force.} 
The growth of science is readily illustrated by the numbers of faculty members and  faculty-in-training. In the United States (US), a country with an established public funding system, this growth is largely driven by federal funding initiatives.  As a result, the scientific endeavor, in particular the scientific labor force, is sensitive to sudden policy shifts, such as the NIH budget doubling that occurred over the 5-year period 1998-2003, the 2009 federal stimulus plan, and the  subsequent 2013 budget sequestration in the US  \citep{NIHBudgetdoubling,EconScience,SystemicScienceFlaws,ScienceFunding}. Other countries are also susceptible to volatile funding, as was the case with austerity measures in  European countries following the recent global recession.

To illustrate  the growth  of the scientific population we have analyzed  the number of US  graduate students, postdoctoral fellows, and faculty members over a 40-year span for the natural and health sciences \citep{D0,D1,D2}. Fig. \ref{f0}A shows that the number of NSF and NIH funded graduate students and postdocs is growing at roughy $2.2\%$ and $3.7\%$ annual rate, respectively. For comparison, these growth rates are slightly larger than the growth rates of the global population over the same period, which according to the US Census Bureau is between $1\%$ to $2\%$ \citep{GlobalPop}. 
Fig. \ref{f0}(B-D) show the growth in the size $S(t)$ of the postdoctoral and faculty population in six fields, with respect to the base year 1973.  When disaggregated by field, the growth in the academic population no longer exhibits a smooth trend, as in Fig. \ref{f0}A, but instead reflects the nuances of federal steering. Notably, the scale of the growth factor is significantly larger for the postdocs than for the tenure-track faculty, reflecting the formation of a bottleneck in the career pipeline  \citep{EconScience,Bottleneck,PhdFactory}.  
Indeed, Figs. \ref{f0}E and \ref{f0}F show how the overall ratio of faculty to postdocs, an indicator of promotion likelihood,  has significantly decreased over the last 40 years. 

These trends reflect the ways in which the academic profession is growing. The embedding of scientists into large teams is a corollary of this growth. Little is known about how these trends are impacting the levels of competition and career sustainability, but there are signs of potential problems \citep{SystemicScienceFlaws,ChasingMoney}. In particular, as we shall demonstrate in the next section, the new entrants into the scientific community swell the number and size of teams. As a result, the important mentor-mentee relation may be at risk; in addition to the ethical conflicts that arise when a mentor has more than one mentee \citep{MentoringEthics},  mentors have less time per mentee due to time constraints. This unbalancing trend in the mentor-to-mentee ratio  may negatively impact the graduate training experience by reducing the opportunities for mentors to offer  psychosocial help \citep{GradSchoolMentoring} and to develop strategies to decrease undesirable behavior  \citep{YouthMentoring}.  Furthermore,  it may adversely affect a broad range of mentoring outcomes \citep{MentoredNonMentored,FacultyMentoring}, including the development of academic identity and  academic expectations \citep{NextGenFaculty},  and in general, the  cross-generational cultivation of values. \\

\noindent{\bf Team size growth.} 
As the complexity of research projects increases, collaboration within teams becomes a key feature of  the science system. To illustrate the steady growth of team sizes, we have undertaken a historical analysis of coauthorship in scientific  publications and coinventorship in patents. The public domain teams recorded in scientific publications  vs. the private domain teams captured by patent applications  offer a comparative perspective on the role of teams in  R\&D during the last half-century.

Specifically, we analyzed coauthorship patterns in four Thomson Reuters Web of Knowledge (WOK) publication datasets: (i) the  biology journal Cell, (ii) an agglomeration of 14 high-impact economics journals,   (iii) the New England Journal of Medicine (NEJM), and (iv) the Physical Review Letters (PRL). 
These journals represent four distinct research domains, chosen to demonstrate that the pattern of exponential growth is common across the datasets analyzed. The discipline-specific growth rates likely reflect the differences in the production of knowledge within each discipline. We refer the curious reader to \citep{TeamScience} for a broader subfield analysis, which includes Arts \& Humanities,  and it also discusses the relative citation impact premium attributable to teams.
 
 \begin{figure*}
\begin{center}
\includegraphics[width=0.75\textwidth]{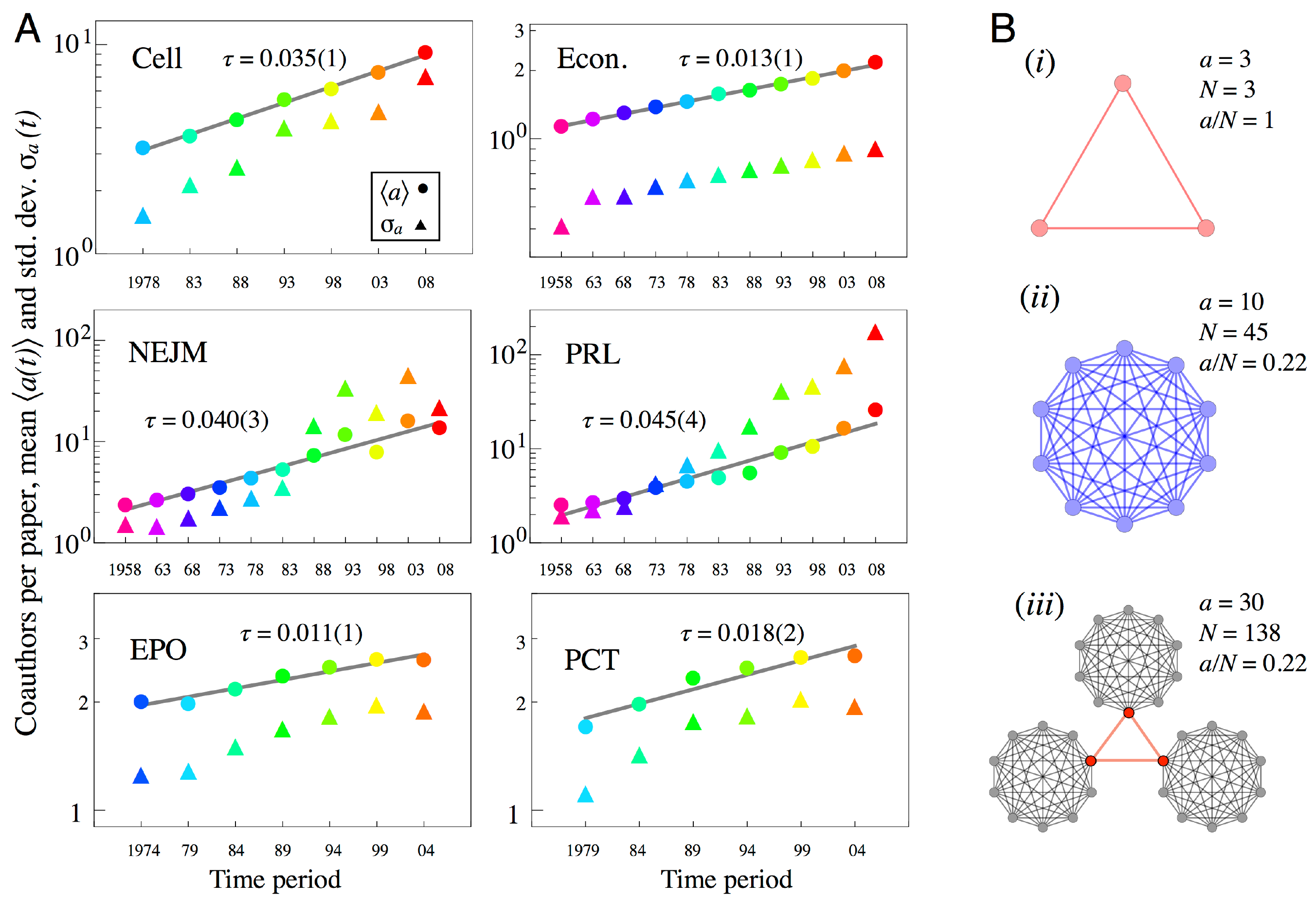}
\caption{\label{f2}  {\bf Persistent growth of team size and the increasing dilemma of sharing credit.}  (A) For each 5-year period we plot the mean of the distribution $\langle a \rangle$ and the standard deviation $\sigma_{a}$, and report the annual growth rate $\tau$ calculated for $\langle a(t) \rangle$. Years listed are the start year for each  of the 5-year non-overlapping periods. (B) There is an increasing complexity with team size; $a$ denotes the number of team members (nodes), $N$ the number of ``associations'' (links), and so the ratio $T\equiv a/N$ is a simple measure for the transparency of the team's activities.}
\end{center}
\end{figure*}

In each dataset summarized in Table 1 we  count for each publication or patent the number $a$ of coauthors (coinventors), a measure that is a proxy for team size.  
 To identify the evolution of coauthorship (coinventorship) patterns, we separated the data into non-overlapping  periods and calculated the complementary cumulative distribution $P_{\geq}(a)$ for each dataset. To put it another way, the value $100 \times P_{\geq}(a)$ indicates the percentage of papers (patents)  that have at least $a$ coauthors (coinventors).
 
Fig. \ref{f1}   illustrates the evolution of $P_{\geq}(a)$ for all six data sets. Please note that the range is long, spanning from unity ($10^{0}$) to more than 1000 ($10^{3}$) on the x-axis (showing team size) and from unity to 1 part per million ($10^{-6}$) on the y-axis (showing frequency).
Hence, we use base-10 logarithmic scale to express the entire range. 
 The key characteristic of these distributions is the persistent shifting towards larger $a$ values over time, indicating the increasing frequency of large teams. This shifting is  becoming increasingly  right-skewed, having a distinct ``extremely large team'' component class that is  emerging in the right tail (evident in the NEJM, PRL, EPO, and PCT datasets for $a>50$). 
 
 Fig. \ref{f2}A shows the growth of the distribution mean $\langle a \rangle$ and  standard deviation $\sigma_{a}$ for each time interval. Note that the time-period color legends are consistent across all panels in Fig. \ref{f1} and Fig. \ref{f2}A, facilitating comparisons. From each time series $\langle a_{j}(t) \rangle$  we estimated the exponential growth rate $\tau_{j}$ for dataset $j$, using ordinary least squares regression of  $\log \langle a_{j}(t) \rangle$. Fig. \ref{f2}A depicts persistent long-term exponential trends for $\langle a_{j}(t) \rangle$, quantified by annual growth rates $\tau_{j}$ in the range of $0.011$ - $0.045$ (see Table 1).
 
It remains to be determined how much of the  growth in team size is produced by social versus  technological  change, and whether the variation in $\tau_{j}$ across disciplines reflects specific socio-economic factors such as size, subfield population composition and population growth, or idiosyncratic publication and funding norms  \citep{TeamScience}. Furthermore, it is unclear if  the recent shift represents a transient reorganization from one regime to another, or if the trend will continue to persist long-term. It is also worth noting that the distribution of team size does not necessarily depend on the sustained growth of scientific production, but instead reflects the relative prevalence of large teams with respect to small teams across a widening range. To illustrate this point, consider the medical journal NEJM,  for which there has been a dramatic  decrease  in the fraction of publications that have a single author - from 21\% in 1958-1962   to 7\%  in  2008-2012. Redistributing this 14\% difference in the frequency of singleton publications, across the entire range of the NEJM distribution, and taking into account that  large clinical trial publications can have in  excess of 700 coauthors,  accounts for a significant portion of the growth in   $\langle a_{j} \rangle$.

Despite these caveats, given these relatively stable trends, it is tempting to  make a crude forecast for the next generation of scientists. If we take the growth trend observed for the journal Cell over the past 35 years (representing a career length), and extrapolate the trend over the next 35 years to 2050, we predict the mean  team size $\langle a(2050) \rangle$  to be approximately 34 coauthors per paper. A similar extrapolation for the EPO growth trend suggests that by 2050 the mean patent will have approximately 4.2 coinventors; for comparison, this is the same as the mean coauthorship for Cell in 1988. For PRL and NEJM the predictions for $\langle a(2050) \rangle$ are significantly greater, being 105 and 74 coauthors, respectively. Overall, these basic trends demonstrate the systemic shifts arising from slow but steady exponential growth in the course of one generation.  While it is unrealistic to expect these trends to extend indefinitely, there remains plenty of room for team size growth, especially considering that the distributions of team size are  heavily right-skewed and so the mean can be dramatically affected by just a few extreme  events. For example, consider the new opportunities in science provided by the ability to obtain crowdsourced research input across the entire population. As such, the upper limit to the number of  participants in a research project may be bounded only by the human population size, as it is evidenced in a recent open laboratory project with roughly 37,000 acknowledged participants \citep{openlabscience}. 

Interestingly,  for the  medicine and physics journals analyzed here, there is a  crossover period, where the standard deviation (measuring the characteristic deviation from the mean value) becomes greater than the mean value, $\sigma_{a}(t) > \langle a(t) \rangle$. This ``tipping point'' marks  the entrenchment of large team science in these disciplines \citep{LittleBigScience}. In the  PRL data this crossover occurred in the 1970s, whereas for NEJM this occurred  in the 1990s.  It is well documented that large team endeavors have existed in physics since the Manhattan project \citep{Rhodes:2012}. Our quantitative analysis points to similar shifts in medicine related to large clinical trials \citep{OASIS5}. Recently, this pattern has been spreading to biology due to large genome projects such as ENCODE \citep{ENCODE}. The ``large team science'' feature has not yet appeared in either the  economics or the patent datasets, although one is left to speculate that it is only a matter of time as long as the right incentives to collaborate are present.$^{2}$\\
 \footnotetext[2]{It is important to note that the intellectual property rights associated with a patent are also shared across all $a$ coapplicants (coinventors  and/or coassignees). Because patenting is based upon proof-of-principle and not necessarily implementation, the commercial rights only need belong  to the person(s) who originated the idea. Furthermore, due to the possibility of direct financial benefits attached to the patent rights, there is a tendency to keep coapplicant  lists from reaching extreme sizes.  Since only the idea is necessary, and prospects of large financial reward are understood, industries encourage patenting ideas almost as quickly as they are generated. And, because most ideas are never implemented, there is little incentive to include people with potential downstream contributions (e.g., those who eventually would implement the idea and/or test it). These reasons account for the significantly smaller team sizes and growth rates in patents with respect to scientific publications. Nevertheless, recent   policies in companies and academic and government institutions   requiring the pre-assignment of an employee's future intellectual property to the employer may be responsible for a systematic shift away from single-applicant patents.}

\noindent{\bf Hierarchy and transparency in large teams.}
The growth in characteristic team size persists over time for each dataset analyzed, and largely reflects the increasing complexity of scientific endeavors. This increasing complexity is also manifest in the organization of scientific teams. Ideally, team leaders efficiently implement a division of labor according to various levels of specialization, so that resources are optimally utilized within the team.

The schematic in Fig. \ref{f2}B demonstrates how the overwhelming number of dependencies between team members in large teams calls for a modular management strategy, which is effected by a hierarchical distinction between team members. Indeed, the maximum number of (undirected) dependencies $N$ in a team of size $a$ is given by $N= a(a-1)/2$. These dependencies (links) represent the multitude of associations between team members. In line with common intuition, the ability  of any given team member to monitor all aspects of the team's operations -- i.e., the ``transparency'' of the operation -- decreases as the mean number of links per person in the team,  $\langle k \rangle = 2N/a$, increases. Hence, for highly connected team networks, the transparency $T\equiv a/N$ decreases significantly with increasing $a$, reaching a minimum value $T_{0}= 2/(a-1)$ for a completely interdependent  team.  

Fig. \ref{f2}B illustrates how the team structure positively affects transparency. By going from team $(i)$ to team $(ii)$ where the team size increases  from 3 to 10, the transparency value decreases by a factor of 5, going  from 1 to 0.2. However, a modular team structure, as demonstrated in team $(iii)$, can overcome the transparency reduction problem. In this case, even though the team size increases by a factor of 3 from team $(ii)$ to team $(iii)$, the transparency value $a/N$ remains the same. Introducing this organizational complexity, however, means that three  team members (red)  are distinguished from the other team members, forming a leadership hierarchy. 

The larger implications of the transparency problem are that for large teams it is difficult  ex post facto to allocate credit  \citep{CreditTaxonomy}, to  assign blame, to justify inclusion or exclusion from coauthor lists,  to disentangle conflicts of interest, and from a practical perspective, to maintain team efficiency \citep{GrowthCareers}. 

\begin{figure*}
\begin{center}
\includegraphics[width=0.8\textwidth]{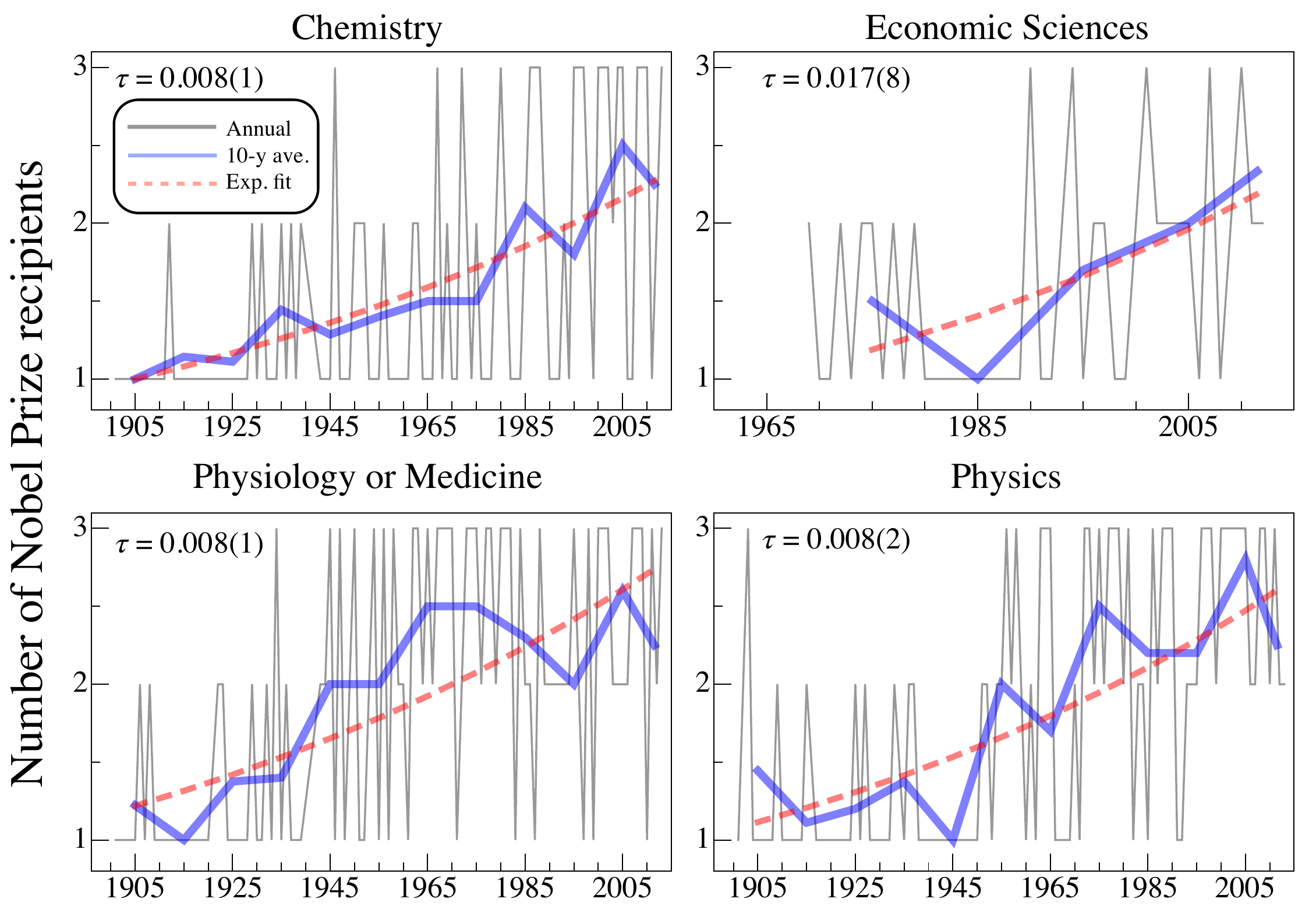}
\caption{\label{f4}  {\bf Increasing cardinality of the Nobel Prize.} The number of Nobel prize  recipients per award (black data), averaged over each decade (blue data), shows steady growth (dashed red curve is the exponential fit of 10-year mean with growth rate $\tau$ indicated in each panel). We provide the estimated growth rate purely as a comparative value between disciplines without implying that the cardinality of the Nobel prize will continue to grow.}
\end{center}
\end{figure*}

These basic conjectures are consistent with recent research on team formation aimed at explaining why some teams far outperform other teams \citep{BuildingGreatTeams}. This research indicates that three team member properties - energy, engagement, and exploration - are crucial factors underlying successful teams. It is also found that structural features, such as  hierarchical ``teams within teams'', can reduce the cohesive engagement among less active team members. Hence, in order to overcome this negative feature of hierarchy, as well as to overcome the transparency problem so that members are aware of each other's contributions, teams must actively focus on high levels of engagement.

In science, the fair allocation of credit is especially relevant in the context of lifetime achievements, which come in the form of career awards and membership in prestigious academic societies. The problem is that career awards such as the Nobel prize, which are limited to a maximum of 3 recipients per award,  can significantly disregard the success that is attributable to  collaboration. Indeed, it is  becoming evident in the Nobel prize award cardinality patterns that the institutions of singleton awards are reaching their limits.  Fig. \ref{f4} shows the number of recipients per Nobel prize award for each of the four science categories \citep{NobelData}. To provide a crude estimate of recent growth rates for comparison, we estimated an exponential growth trend using the 10-year mean calculated within non-overlapping 10-year periods. The  growth trends suggest that an amendment to the 3-person cap on the number of recipients per award should be made for both the ``Physiology or Medicine'' and ``Physics'' prizes, which appear to be outgrowing the upper limit established in the era of singleton science. This example serves as additional  evidence that the incentive system in science is not adapting to the systemic shifts that have occurred alongside the basic growth of the scientific industry. \\

\noindent{\bf Internationalization of scientific networks.} 
The internationalization of global R\&D  reflects the drive to produce high quality output through optimal combination of experts, independent of locality. Trends in cross-border collaboration intensity can indicate the role of distance and geopolitics, 
factors of great relevance for the integration of interdependent innovation systems, e.g., within  the European Research Area \citep{ScienceEurope}. Scientific publication data  provides a good proxy for cross-border activities, yielding insights into various collaboration network properties and the relation between a country's international collaboration intensity, spending per researcher,  and the mean citation impact per paper \citep{SantoWorldCollab}. 

While  the number of publications has been growing steadily, in large and small R\&D systems alike,  it is not well understood at which rate smaller countries are joining the   network of established R\&D systems. To illustrate this integration  process, we analyzed a NSF   database of 264,431 Science \& Engineering publications sampled from the years 1995 and 2010 \citep{D3}. From the counts of the total number of publications  $M_{ij}$  coauthored by country $i$ and $j$, we define  the relative share of country $j$ in the collaboration portfolio of country $i$ in year $t$ as $S_{ij}(t)\equiv M_{ij}(t)/ \sum_{i}M_{ij}(t)$. The relative integration index $g_{ij}(t,\Delta t) \equiv \log [S_{ij}(t)/S_{ij}(t-\Delta t)]=\log [S_{ij}(t)] - \log[S_{ij}(t-\Delta t)]$ measures the relative growth of county $j$ within the portfolio of country $i$ over a given time period $\Delta t$.  The natural logarithm is used so that $S_{ij}(t)$ reflects a relative (percent) change over time, since the value $M_{ij}(t)$ can vary dramatically across the set of countries analyzed.

\begin{figure*}
\centering{\includegraphics[width=1.0\textwidth]{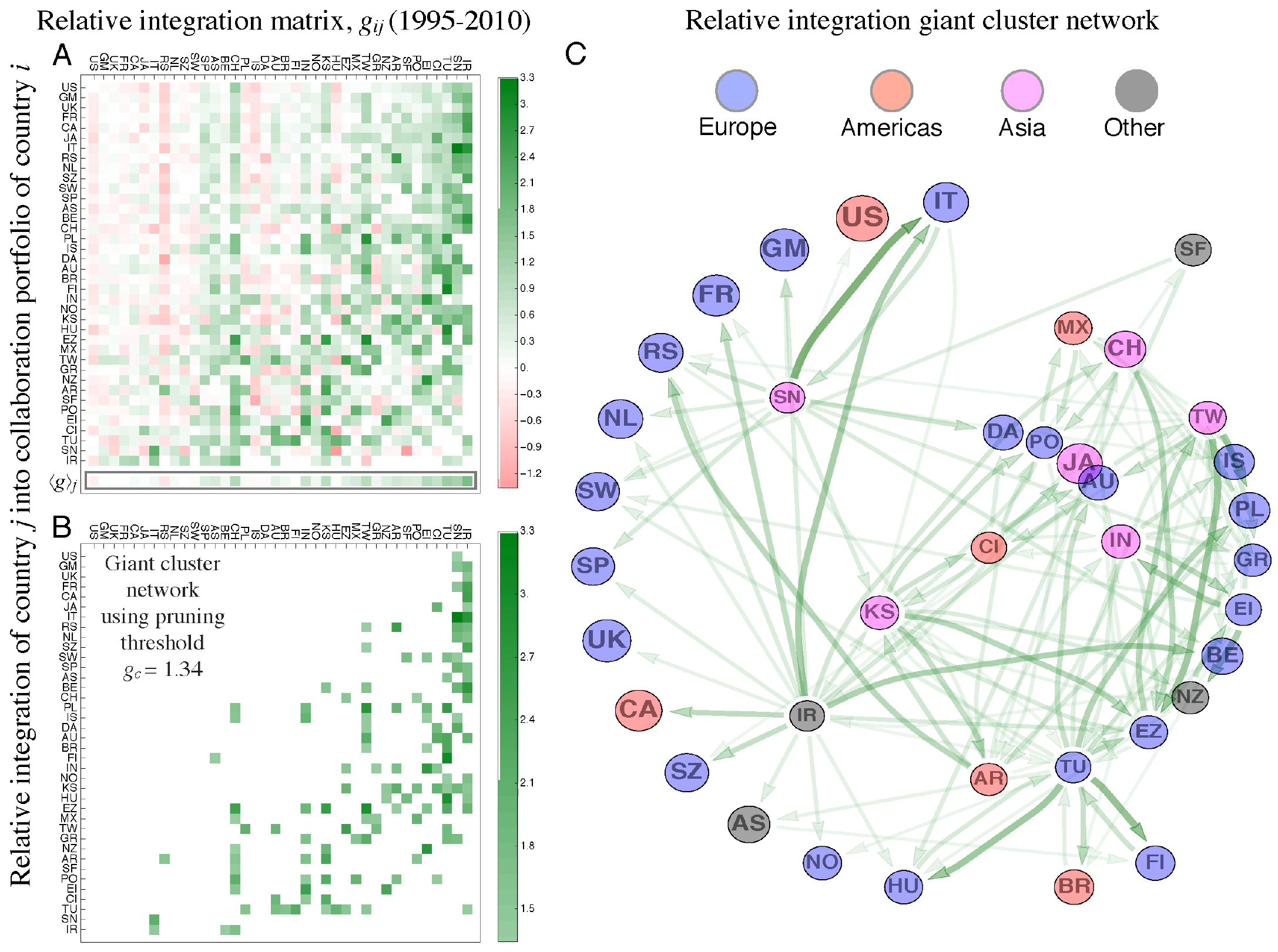}}
\caption{\label{f5}  {\bf Visualizing cross-border collaboration growth.} (A) The relative integration matrix $g_{ij}$ measures how much a country in column $j$ has increased (green) or decreased (red) its collaboration in scientific publications with the other countries (row $i$). The final row indicates the mean value of $g_{ij}$ for each column. China (CN) stands out as a large producer of scientific publications, which has also increased its collaboration share with almost every country shown. (B) By  eliminating (pruning) all matrix values with $g_{ij} < 1.34$ we obtain the minimal spanning cluster. (C) Network representation of the minimum spanning cluster; nodes are countries colored by region with size proportional to   $\log A_{i}(2010)$, and links have thickness and shading proportional to $g_{ij}(1995-2010)$.}
\end{figure*}

Fig. \ref{f5}A shows the 15-year growth  matrix $g_{ij}$ for the top 38 internationally collaborating countries in 2010, where the countries are listed in decreasing order of total publications $A_{i}(2010)$. The mean value $\langle g \rangle_{j} = 38^{-1} \sum_{i} g_{ij}$ of a given country, shown in the bottom row of the relative integration matrix, indicates how much the country is integrating globally; green shading corresponds to positive growth while red shading corresponds to negative growth. The leaders over the 15-year period are Singapore (SN), Iran (IR) and China (CH). In contrast,  Russia (RS) stands out as the only country with a negative mean integration rate (see the Methods section for a full list of country names). The overall trend is for the countries with smaller $A_{i}$ (countries further to the right in the growth matrix) to have the largest integration rates, indicating their recent entry into the global R\&D economy.

Since the relative integration matrix represents the integration between country pairs, it is also insightful to visualize the  systemic correlations that are contained in the network topology. Of particular interest in the context of our analysis are the  sparsely distributed  $g_{ij}$ values that are significantly positive (dark green), which likely represent new country-country links.
In order to visualize the network defined by these relatively large  $g_{ij}$ values, Fig.  \ref{f5}B shows the integration matrix after we eliminate all values with $g_{ij} <1.34$. We choose the pruning threshold $g_{c}\equiv 1.34$ because this is the largest value for which all nodes (countries) in the network are still connected within a single network. In the network science terminology, this minimal set of links that connects the entire system defines what is called the ``giant spanning cluster''$^{3}$.\footnotetext[3]{Interestingly, the United States is the first country to be eliminated from the giant spanning cluster network for $g_{c}>1.34$. This feature follows from the fact that the US has long been a collaboration hub, having already a large $M_{ij}(1995)$ for all the countries shown. Hence, the percent growth of counties within the US portfolio is relatively  small due to  upper limits in the amount that collaboration can increase. Nevertheless, China, Singapore, Turkey, and Iran show signs of significant integration with US research over the 15 year period.} Fig.  \ref{f5}C illustrates the giant spanning cluster network and  highlights the role of incoming partners on the globalization of science. The global integration hubs  Iran (IR), Singapore (SN),  Turkey (TU), China (CN), and South Korea (KS) are characterized by the disproportionate number of integration links  connecting them to countries from every other geographic region represented in the data.

Hence, how does globalization affect issues of team ethics? 
While it is straightforward to argue that the integration of the global R\&D economy is  good for both science and society, in the following discussion  we highlight practical dilemmas arising from  significant differences  between the social norms and ethical boundaries framing scientific activities in different countries.  We argue that as countries with different economic levels, language, governance, laws, and  cultural teamwork norms, continue to integrate their R\&D programs, it will become increasingly difficult to disentangle subjective ethical considerations.  As a result, we predict that ethical conflicts of this type will  become  increasingly prevalent.

\section*{Discussion}
Here we address the intersection  of our quantitative analyses  with broad themes pertaining to team science  ethics. The persistent trends in the skewed growth of the scientific labor force, the increasing organizational complexity arising from increasing team sizes, and the global increase in the intensity of cross-border collaboration activities, have consequences for at least  six  issues of ethics outlined below:\\

\noindent {\bf1. Ethics of credit: Who to reward and who to blame?} \\
\noindent The assignment of credit is fundamental to the reward scheme in science \citep{StephanJEL, OpenScience, EconScience}. The basic staple of the science credit system is the credit associated with a publication. This is supposed to be shared across all $a$ coauthors independent of their rank. The problem is that large teams in science have a pyramidal structure, with the scientists who masterminded the project and obtained the funding at the top. These project leaders reap co-authorship gains from the entire hierarchy below them. For example, it is not uncommon for directors of large particle physics laboratories to publish upwards of 50 papers in a good year. Hence, due to the nonlinearity of the underlying hierarchy and the broad range of team sizes in science (Fig. \ref{f1}), even fractionally distributing publication and citation counts  could introduce unfair bias.

In another hierarchical corps, the military, a scheme of  battle badges  evolved in order to  distribute credit among the team members contributing towards large scale goals. The science corps has developed a similar system of allocating special ``badges'' to the first author and the corresponding author; the latter typically being the principal investigator who led the project. However, variations in the norms of determining authorship order do exist across disciplines.$^{4}$  \footnotetext[4]{In the natural sciences,  the first and corresponding author(s) are typically distinguished from other coauthors. In economics alphabetical ordering of the coauthor list is often the norm, thus eliminating special credit for the lead author and principal investigator. Furthermore, in economics it is common that graduate student data collectors and data cleaners are not included in the coauthor list and only acknowledged in a footnote.} 

Recently, the methods for distinction have  begun to change in prestigious journals, allowing authors to designate their particular roles, e.g., designed research, performed research, contributed new reagents/analytic tools, analyzed data, or wrote the paper. This shift reflects the need for scientists within teams to distinguish themselves \citep{CreditTaxonomy}. Being associated by name with a seminal paper can be a major career boost, 
especially for the first and corresponding authors. Yet, as the team size increases, possibly across multiple groups and hence across multiple principal investigators, a great difficulty in selecting first and corresponding authors arises \citep{FirstAuthorNature}. One resolution to the problem is the practice of multiple publications, whereby several variations of the same paper are submitted to various conferences and journals with permuted author list orderings  \citep{CoauthorshipEthics}. This practice, however, contradicts the system of precedence and is considered in some disciplines as ethical misconduct.

The problem of credit attribution among coauthors is not new. Historically, even in the case of small teams, sometimes the contributions of junior scientists are unfairly allocated to the  senior scientist. However, the possibility of  unfair reward  increases with team size for two main reasons. First, because it becomes increasingly difficult to discriminate efforts of the individual participants as demonstrated in our discussion of the team transparency $T$. And second, because  it becomes increasingly difficult to discriminate who should and who should not be included as a coauthor. While there have been recent  efforts to develop quantitative  methods that factor in team size in allocating publication and citation counts  \citep{ScientistMetrics,CollabMetrics}, accounting for variations in team organization and specialization remains a core issue in the fair distribution of scientific credit. 

The former considerations address  scenarios where science goes well. The converse scenario  raises the issue of  who to blame  when science goes wrong. Questionable tactics pervade everyday  scientific practice, including several that are particularly relevant to team settings,  such as  failing to acknowledge credit for research ideas, misallocation of authorship credit, multiple publications, and non-disclosure of conflicts of interest  \citep{BehavingBadly}.  Retraction of scientific papers is quite common, with roughly 2/3 of retracted papers related to misconduct  \citep{retraction}. Not all coauthors, however, may agree with the retraction, which further complicates matters. By way of example, recent claims of faster-than-light neutrinos in a large team setting resulted in a subsequent retraction. However, a fraction of the team including the principal investigator insisted on the validity of the finding despite mounting evidence  that the initial results were flawed by experimental error. In the case of retraction due to experimental error it may be difficult to trace the blame to any single individual. In the case of retraction due to fraud (for example, the Woo Suk Hwang controversy)  the blame may be entirely  attributable to the principal investigator whose individual actions can compromise  the  efforts, reputation, time, and careers of each team member  \citep{NatureEthicsFraud,retractionimpact}.

Beyond the scientist's responsibility towards her/his team, lies the scientist's responsibility towards society. It was in the early 20th century when certain risks to humanity from scientific progress became evident and moral questions about the role of individual scientists were raised. Such was the case of Fritz Haber (Nobel Prize in Chemistry, 1917), who discovered a method to synthesize ammonia with applications in fertilizers and  chemical bombs. By the mid 20th century, Oppenheimer became the tragic figurehead of the  Manhattan project, which epitomizes the  dilemma associated with the moral responsibility of individual scientists embedded in larger socio-political programs. With increasing numbers of large projects faced with ethical dilemmas,  and with many of these projects having multiple figureheads and a hierarchical structure that tends to cloud the channels of responsibility \citep{Schonmisconduct},  paradoxically moral responsibility has been shifting towards the scientific commons. \\

\noindent {\bf2. Parasitic authorship.}\\
In a large team setting, it is difficult not only to determine coauthor order, but also to determine who merits inclusion in the coauthor list.  After all, the addition of a single coauthor, from $a \rightarrow a+1$, appears to be only a marginal modification when $a$ is large. Some senior researchers take advantage of this coauthorship culture by exploiting uncertainties or ambiguities in research guidelines and thus prospering in poorly regulated, grey areas  \citep{WhiteBull}. To limit this problem, scientific institutions need to better define and impose ethical codes for authorship credit, materially discouraging free-riding  and other corrupt authorship practices, such as bartering for coauthorship \citep{ChinaCouathorshipBazaar}.

The central question is what constitutes coauthorship? The criteria differ among disciplines and may be journal dependent. Even within a given community, there may not be consensus on the criteria that constitute {\it significant} contributions meriting coauthorship \citep{WhiteBull}. By way of example, with English becoming the de facto language for science, many international teams must include members solely for the purpose of helping and reviewing in the writing process \citep{CoauthorEthics2}. But does this constitute authorship, or does it fall under the category of  support? Many would argue that in the case of support, the contribution should only be mentioned in the acknowledgement section of the manuscript. But is this a fair way of rewarding a crucial feature of scientific discourse and refinement \citep{peerhelp}?\\

\noindent {\bf3. Conflict of interest.}\\
Many ethical dilemmas in science arise from a conflict of interest that may emerge between individual scientists but can also be made manifest between the scientist and the greater scientific community. For example, self-interest and favoritism are known to  undermine the publication review process, which relies on individuals to treat each other fairly, sometimes  in light of undisclosed competing interests. For this reason it is widely accepted that previous coauthors or mentor-mentee pairs should not be allowed to review each other's manuscripts \citep{PeerReview}. It has also been proposed that mentors with more than one mentee are implicitly incapable of meeting the  ethical obligations of a mentor \citep{MentoringEthics}. Such conflicts of interest between individuals  become more likely as  team sizes grow and the interconnections in the ``invisible college''  become  unavoidable. 
Furthermore, as the scientific enterprise expands and competition for limited resources increases (sometimes even within the same team) the risk-to-return tradeoff  may incentivize unethical success strategies.  In this respect, a conflict of interest between the scientist and the scientific commons emerges, whereby bad behavior may evolve as individuals reconsider their identity and responsibilities  within the scientific system
\citep{BehavingBadly,Schonmisconduct,AcademicBirthRate,ScientificMisconduct,SelfInterestResearchBehavior}.  

The level of competition in science can be readily illustrated by considering   the number of successful NIH R01 grants and grant-holders  relative to the number of submissions  and the total size of the applicant pool. Along those lines, a recent  news focus in the journal Science puts into perspective the decreasing total NIH budget from 2003 - 2014 and how it has impacted various actors in biomedical science \citep{ChasingMoney}. Their  numbers show that the NIH budget decreased from its peak at roughly 22 billion USD in 2003 to  17 billion USD  in 2014 (values deflated to 1998 USD); meanwhile, the success rate of R01 grants has halved while the number of {\it funded}$^{5}$ principal investigators has increased by  5\% from 2000 to 2013. The report goes on to show that  the average scientist's age at the time of his/her first R01 grant has increased from 36 in 1980 to 42 in 2013, and likewise, that that the percentage of principal investigators over 65 has increased from 3.5\% in 2000 to 7\% in  2010.  Aside from the unsustainable generational economics of science, these trends also indicate 
that the young scientists are assuming a disproportionate amount of the financial burden, likely due to the granting system and other features of science careers, which are based upon the principles of cumulative advantage.  \footnotetext[5]{These numbers reflect the number of funded individuals, and thus do not account for  the unfunded population. By inverting the success rates we can estimate that the unfunded population has increased by 22\% over the same period. This scenario is further exacerbated by the fact that a small number of principal investigators (6\% of senior scientists) receive a disproportionate amount of the annual funding (28\%) provided through NIH grants \citep{ChasingMoney}.}

In a system where less funding must support a growing population of scientists, one quick solution is to fund teams instead of individuals, a step up from the Howard Hughes Medical Institute's (HHMI) mission to fund ``people not projects''. For the moment, there is tension between the de-facto tenure requirement in biomedical departments  that an assistant professor must obtain an R01 grant and  the grant competition levels that render this prospect a statistical impossibility, even for stellar young scholars. To address this problem,  funding needs to be  increased,  the number of scientists in the pipeline needs to decrease \citep{SystemicScienceFlaws}, and specific to the case for tenure, the  criteria for research scholarship  need to be adjusted to better reflect contributions in team settings.

The categories  for documenting teaching, research, and service activities in the tenure process (see \citep{TenureBook} pp. 49--52) are numerous and can vary from institute to institute. The documentation of teaching activities is more transparent, as teaching hours and student evaluations of the candidate are relatively easy to evaluate ex post facto. However, determining an individual's contribution to research scholarship in an ex post facto evaluation can be extremely difficult if the output of journals, books, grants, patents and presentations are complicated by a variable and non-negligible team size. 
Moreover,  while it is important {\it not} to discount the value of authorship in multiple-author research products, so that tenure productivity criteria aren't biased against team-oriented scientists, it is also important {\it not} to discount the value of  investigator status in {\it Multiple Principal Investigator} (multi-PD/PI) grants. Concerning the service component of tenure criteria, there is a growing consensus calling for the acknowledgement of  patenting and other commercially valuable activities, which have a ``community service'' component \citep{TenureRevision}.

\noindent {\bf4. Mentor-mentee ethics.}\\
The incentives to publish (or not publish) for a young scientist are different than those of an established scientist. For example, what happens in the case that a mentee's  findings are in disagreement with the previously published findings of the  mentor? Further mentor-mentee issues may arise in large teams  where the ratio of mentors to mentees is small (as shown in Fig. \ref{f0}). In these cases, the mentor may be unable to guide each student or postdoc individually due to time constraints. As a result, the benefits of mentorship become  diluted with negative implications for academic character building which is the basis of virtue ethics \citep{Schonmisconduct}.

An additional issue that tests the mentor-mentee relationship is the narrowing bottleneck in academia  \citep{AcademicBirthRate,Bottleneck}, whereby an increasing number of Ph.D.s and postdocs are being churned by large multi-institutional project grants that likely have a weak impact on the number of new tenure-track openings. As the prospects of climbing the career ladder in Academia are often overstated, with the career outcomes  traditionally being poorly documented \citep{EconScience,StephanPhDBubble}, many young scientists have likely been ``lured''  into postdoctoral traps within large projects. This raises the question: Are the next crop of scientists trained to be leaders or to just fit into a large production line? And once they enter the tenure track, do the lessons they learned in their ascent reflect positive scientific values? Or do they reflect a system engaged in productivity at the expense of quality, the choice of conservative research projects over innovative risk-taking ones,  and  pathologically competitive attitudes that run counter to socially beneficial progress \citep{ScientificMisconduct,CareerCompetition,CompetitionEffects}? \\

\noindent {\bf5. International variations in ethics codes.}\\
The norms of leadership, management, and promotion can be largely country dependent \citep{EuropeTenureSystems}. Moreover, the norms for ethical conduct in science \citep{crossculturalvariation} and the laws reflecting bioethical standards on research topics involving stem cells \citep{lawvariation},  experiments with animals, and human clinical trials, can also vary significantly across counties  \citep{EthicsofCrossborderColl}.

For example, the localization of proprietary  biomedical R\&D in countries with less restrictive stem-cell bioethics legislature \citep{USEUstemcellethics, hespublications} reflects how these  variations across countries have entered into corporate strategy. 
While it may be in violation of local ethics codes and legislature to work on specific types of stem cells in one country, should it also be in violation  to collaborate with partners in another country that does allow the controversial stem cell line? To give a  specific example, the outsourcing of clinical trials to poor regions of India  has become a popular method of side-stepping local ethical and economic impediments  \citep{CaseIndia,OutsourcingClinicals}. 

 Fig. \ref{f5} shows that in the era of large team science, more collaborations are crossing national borders involving developing economies and possibly third world poverty. Adhering to  local ethics codes in a global system  is important for the building of character and identity. To facilitate the decision making process when international teams encounter conflicts in local ethics codes,  the global standardization of ethical norms is crucial \citep{worldreport1,worldreport2}.\\

\noindent {\bf6. Universality of norms.}\\
We have already mentioned how international standards can vary significantly. Another relevant question is whether we  should expect for the ethics of small team science to map across scale and apply unflinchingly to large team science.  Several features of large team science challenge the institutions constructed for small team science, namely the reproducibility of such large projects (inherently requiring complementary large teams committed to verification), and the distribution of credit to all participants. Finally,  increasing team size  is also accompanied by the growth of interdisciplinary science: Is it possible to expect that social norms of ethical publication conduct be shared across disciplines?

\section*{Conclusion}
We have used quantitative analysis to document trends in scientific operations that bear ethical ramifications and call for introspection and open discussion.
Over time these trends will affect an increasing fraction of  scientists, whose careers will depend on team activity. Even within the social sciences, where historically team sizes have been small, the trends reveal slow but persistent growth. For example, our analysis indicates that by the year 2050 the mean publication  team size in economics will be $\langle a(2050) \rangle \approx 3.5$ coauthors, which is comparable to the mean publication team size in cell biology during the 1980s and in physics during the 1970s. Moreover, in the next 35 years -- typifying a scientific generation --  we also project that 5\% of the teams will be greater than 100 coauthors  in physics and greater than 50 coauthors in cell biology. Hence, the ethics issues we have outlined will become increasingly pressing.

The first issue raised is how persistent growth in  team size poses a challenge to  the longstanding credit system in science, and calls into question the appropriateness of singular achievement awards in team settings.  In our discussion, a theme has developed around the heterogeneity of the actors in scientific teams and the distinct role of team leaders who often gain a disproportionate share of credit. 
When this credit bias is coupled with limited upward mobility in the research career ladder, it creates a state of ``haves and have-nots'' that tests scientists' attitudes and behaviors \citep{BehavingBadly}.  

Concerning unethical behavior, further research is needed to investigate how to incentivize cooperation and ethical practice in the team environment, likely calling for new team ethics paradigms. Sanctioning bad behavior in a team environment has benefits, as there is recent evidence that the role of organizational (in)justice, and perceptions thereof, can have an impact on a scientist's identity within the scientific system, and can affect his/her propensity to behave or misbehave \citep{ScienceOrganJustice1,ScienceOrganJustice2}. Furthermore, evidence from organizational game theory suggests that policies that punish unethical behavior should be widely adopted, since institutions with sanctioning are more preferred and offer a competitive advantage  over those without \citep{Freerider}. To this end, it is important to establish  guidelines for sanctioning, both internal and external to specific teams,  which discourage parasitic coauthorship and other bad  behaviors   that are  particular to team settings. Bringing these issues to light may be the first step to establishing a more ethically conscious scientist. However,  providing solutions to the problems raised here will be challenging since monitoring ethical standards and sanctioning misbehavior is  difficult in large team endeavors  due to the transparency problem.

An insidious problem highlighted is how a  large team environment may hinder the  cross-generational transmission of values  from mentor to mentee, undermining the building of virtuous academic characters.  Over time this may lead to gradual erosion of ethical standards across science. To fill the gap, there is need for policies that aim to cultivate morality. Such policies should promote a bottom-up educational approach with emphasis on humanistic values, starting with a student's first introduction to science in secondary school. In a very general sense, cultivation of team science ethical values should become a corollary of the longstanding scientific method.

A body of ethical scientists is indeed an invaluable community resource since the support of social norms is a self-reinforcing process, gaining strength with adoption size. This is a virtuous cycle to which we are likely to fall if we address the emerging team science issues early. The alternative is a vicious cycle that we should aspire to avoid.\\

 \begin{table}
 \centering
 { \small
 \caption{Summary statistics for the journal and patent datasets analyzed. The exponential growth rate $\tau$ (per year) is estimated using ordinary least squares regression; the standard error is enclosed in parentheses. Multiply growth rates by a factor of 100 to obtain the percentage growth.}
\begin{tabular}{@{\vrule height 10.5pt depth4pt  width0pt}lc|c|c}\\
\noalign{
\vskip-11pt}
\vrule depth 6pt width 0pt    &   & Articles /   & Team size \\
\vrule depth 6pt width 0pt \textbf{Dataset}  &  Years & Patents & growth rate $\tau$ \\
\hline  
\hline
Cell &  \  \ 1978 -- 2012 \ \  & \ \ 11,637 \ \ & \ \ 0.035(1) \ \  \\
14 Economics journals &  1958 -- 2012 & 36,466 &  0.013(1) \\
New England J. Medicine &  1958 -- 2012 & 18,347 &   0.040(3) \\
Physical Review Letters  &  1958 -- 2012 & 98,739 &   0.045(4) \\
\hline
\hline
European Patent Office  &  1974 -- 2008 & 2,207,204 & 0.011(1)\\
Patent Cooperation Treaty  &  1979 -- 2008 & 1,695,339 & 0.018(2)\\
\hline
 \hline
\end{tabular}
\label{table:journals}}
\end{table}

\section*{Data \& Methods} 
\noindent{\bf Publication and patent collaboration data.}
Publication data for the journals Cell,  the New England Journal of Medicine (NEJM), Physical Review Letters (PRL), and 14 top economics journals, American Economic Review, Econometrica,  Journal of Political Economy, Journal of Economic Theory, Journal of Econometrics,  Journal of Financial Economics, Journal of Finance, Journal of Economic Growth, Journal of Economic Perspectives, Journal of Economic Literature, Quarterly Journal of Economics, Review of Economic Studies, Review of Financial Studies, Review of Economics and Statistics,  were downloaded  from Thomson Reuters Web of Knowledge for the 55-year period 1958--2012. For the natural science journals we restricted our
analysis to publications denoted as ``Articles", which excludes reviews, letters to editor, corrections, and other content types.  For the economics publications we restricted our analysis to the publication types: ``Articles,'' ``Reviews'' and ``Proceedings Papers''. We obtained the patent data from the Organization for Economic Cooperation and Development (OECD) \citep{OECDPatents}: Years 1974 -- 2008 for European Patent Office (EPO) patents and 1979 -- 2008 for Patent Cooperation Treaty (PCT) patents. We obtained the NSF Science and Engineering Indicators data from \citep{D0,D1,D2,D3}.\\

\noindent{\bf International collaboration network data.} 
Article collaborations are tabulated using a whole-count basis whereby a country is counted only once per paper even if there are multiple affiliations with a given country address. Article data from Thomson Reuters Web of Science covers journals indexed in Science Citation Index and Social Sciences Citation Index.
Country abbreviations are:
United States	(US),
Germany	(GM),
United Kingdom	(UK),
France	(FR),
Canada	(CA),
Japan	(JA),
Italy	(IT),
Russia	(RS),
Netherlands	(NL),
Switzerland	(SZ),
Sweden	(SW),
Spain	(SP),
Australia	(AS),
Belgium	(BE),
China	(CH),
Poland	(PL),
Israel	(IS),
Denmark	(DA),
Austria	(AU),
Brazil	(BR),
Finland	(FI),
India	(IN),
Norway	(NO),
South Korea	(KS),
Hungary	(HU),
Czech Republic	(EZ),
Mexico	(MX),
Taiwan	(TW),
Greece	(GR),
New Zealand	(NZ)
Argentina	(AR),
South Africa	(SF),
Portugal	(PO),
Ireland	(EI),
Chile	 (CI),
Turkey	(TU),
Singapore	 (SN),
Iran	(IR).\\
  
\noindent{\bf Acknowledgements} 
We would like to thank Sarah K. A. Pfatteicher for her helpful feedback. AMP acknowledges support from the IMT Lucca Foundation and support from the Italian PNR project ``CRISIS Lab.'' IP and IS acknowledge support by the National Science Foundation via grant \# 1135357, entitled `EESE-Experiencing Ethics'. Any opinions, findings, and conclusions or recommendations expressed in this paper are those of the authors and do not necessarily reflect the views of the funding agencies.\\


\begin{widetext}


\end{widetext}

\end{document}